\documentclass[journal]{IEEEtran}
\usepackage{subcaption}
\usepackage{mathtools}
\usepackage{amssymb}

\usepackage{booktabs}
\usepackage{comment}
\usepackage{cite}

\ifCLASSINFOpdf
   \usepackage[pdftex]{graphicx}
  \usepackage{xcolor}

   \graphicspath{Fig/}

\else

  \usepackage{here}
  \usepackage[dvips]{graphicx}
  \usepackage[driver]{graphicx}
  \usepackage{epstopdf}
  \AppendGraphicsExtensions{.tif}
  \graphicspath{Fig/}

\fi
\usepackage{graphicx}

\usepackage{tabularx}
\usepackage{amsmath}
\usepackage{amsfonts}
\usepackage{amsthm}
\usepackage{bm}
\usepackage{mathtools}

\usepackage{algorithm,algpseudocode}

\usepackage{enumitem}

\theoremstyle{definition}

\newtheorem{rem}{Remark}

\title{Sparse Channel Estimation in Wideband Systems with Geometric Sequence Decomposition}

\usepackage{graphicx}

\author{Woong-Hee Lee~
        and Ki Won Sung,~\IEEEmembership{Member,~IEEE}
\thanks{This work has been accepted for publication in IEEE Wireless Communications Letters in October 2021.}
\IEEEcompsocitemizethanks{\IEEEcompsocthanksitem  W.-H. Lee is with the Department of Control and Instrumentation Engineering, Korea University, Sejong-si 30019, Republic of Korea (e-mail: woongheelee@korea.ac.kr).
\IEEEcompsocthanksitem K. W. Sung is with the School of Electrical Engineering and Computer Science, KTH Royal Institute of Technology, Stockholm, 164 40, Sweden (e-mail:sungkw@kth.se).}
}

\begin{document}


\maketitle

\begin{abstract}
The sparsity of multipaths in the wideband channel has motivated the use of compressed sensing for channel estimation. In this letter, we propose a different approach to sparse channel estimation. We exploit the fact that $L$ taps of channel impulse response in time domain constitute a non-orthogonal superposition of $L$ geometric sequences in frequency domain. This converts the channel estimation problem into the extraction of the parameters of geometric sequences. Numerical results show that the proposed scheme is superior to existing algorithms in high signal-to-noise ratio (SNR) and large bandwidth conditions.
\end{abstract}

\IEEEpeerreviewmaketitle


\section{Introduction}
In wireless communication systems, wideband channels can be characterized by a few dominant channel tap coefficients \cite{bajwa10}.
This means that the channel impulse response (CIR) exhibits a sparsity in time domain. Therefore, the principle of compressed sensing (CS) \cite{donoho2006compressed} has garnered attention in channel estimation to reduce the pilot overhead of wideband systems. 

One of the popular methods of exploiting the time domain sparsity is to employ orthogonal matching pursuit (OMP) \cite{qi14,mohammadian2016compressive} or its variants \cite{zhang16,wan2018accurate}. However, it has two problems. First, the excess delay of the multipaths appears in the continuous time domain, and therefore there is no guarantee that the CIR occurs exactly on the grid of the sampling interval. Consequently, it is difficult to form a complete sparse vector, only to deal with an approximately sparse vector with energy dispersion. Second, OMP selects the basis with the highest correlation sequentially. Hence, it is vulnerable to inter-path interference coming from the non-orthogonality among multipaths. As a result, the required number of bases must be numerous, which in turn requires high computational complexity. To overcome the limitation of finite dictionaries, methods to find a solution in a continuous space through optimization techniques have been studied \cite{wright2009sparse,tang2013compressed,pejoski2015estimation}. Notably, a method based on atomic norm has been introduced in \cite{tang2013compressed}, and it has been applied to the sparse channel estimation in \cite{pejoski2015estimation}. However, this high-resolution technique also has a condition of the permissible correlation for the superposed signals to be decomposed.

To address the issues of existing sparse channel estimation methods, we take an approach similar to matrix pencil method (MPM) \cite{hua1990matrix} and propose a novel scheme, namely geometric sequence decomposition based sparse channel estimation (GSD-SCE). Assume that a channel consists of $L$ mutipaths. Conventional methods typically divide the grid in time domain as finely as possible and capture $L$ peaks by matching the most correlated bases. Instead, GSD-SCE directly obtains the parameters of $L$ multipaths from pilot estimation. To achieve this, we depart from the fact that each peak in time domain can be viewed as a geometric sequence in frequency domain. This implies that we can think of a CIR as a non-orthogonally superposed $L$ geometric sequences in frequency domain. Therefore, the sparse channel estimation problem is converted into the acquisition of the parameters of the geometric sequences while we can observe only the superposition of the sequences.

The converted problem is efficiently solved owing to the recently developed method of geometric sequence decomposition with $k$-simplexes transform (GSD-ST) \cite{lee2019geometric}. In \cite{lee2019geometric}, sampled radio waves in time domain are described as geometric sequences to distinguish multiple random access requests. This letter explores the potential of the geometric sequential representation in frequency domain to pinpoint multipaths in channel estimation.

\section{System Model}

In an orthogonal frequency-division multiplexing (OFDM) transmitter, a series of $N$ symbols, $\mathbf{x} \in \mathbb{C}^N$, are converted into time-domain signals using the inverse fast Fourier transform (IFFT). Next, a cyclic prefix (CP) is inserted to preserve the orthogonality and to eliminate the inter-symbol interference (ISI).

Let $L$ be the number of multipaths. Obviously, we assume $L \ll N$. The CIR, $h(\tau)$, can be modeled by the sum of Dirac-delta functions as follows:
\begin{equation}
    h(\tau)=\sum_{l=0}^{L-1} \alpha_l \delta(\tau-\tau_l),
\end{equation}
where $\alpha_l$ and $\tau_l$ are the complex channel gain and the excess delay of  $l$-th path, respectively. The channel can be reasonably assumed constant during an OFDM symbol.

At the receiver, the CP is removed before the fast Fourier transform (FFT) process. Assuming ideal synchronization at the receiver, let $\mathbf{s} \in \mathbb{C}^N$ denote the discrete signal obtained by $N$-point FFT after sampling the signal experiencing the channels at the receiver. Then, $\mathbf{s}$ can be represented as follows:
\begin{equation}{\label{ds}}
    \mathbf{s} = \mathbf{x} \odot \mathbf{h} + \mathbf{w} = \{ \mathbf{x}[n]\sum_{l=0}^{L-1} \alpha_l e^{-j2\pi n \Delta f \tau_l} + \mathbf{w}[n]\}_{n=0}^{N-1},
\end{equation}
where $\odot$ denotes the operator of the Hadamard product.
Here, $\mathbf{h} \in \mathbb{C}^N$ is the channel frequency response (CFR), which is the $N$-point FFT of $h(\tau)$, and $\mathbf{w}[n]$ is the additive Gaussian noise with zero mean and variance $\sigma_w^2$ for all $n$.

We assume that $\mathbf{x}$ consists of $P$ equispaced pilot symbols and $(N-P)$ data symbols. Let $\mathbf{x}_T \in \mathbb{C}^P$ denote the sequence of pilot symbols for channel estimation, which can be represented as follows:
\begin{equation}
    \mathbf{x}_T =  \{ \mathbf{x} [pK] \}_{p=0}^{P-1},
\end{equation}
where $K$ is the index spacing of the pilot assignment.
For simplicity, let $\beta$ be the transmitted pilot symbol, i.e., $\mathbf{x}_T [p]=\beta$ for all $p$, which is predetermined. Then, $\mathbf{s}_T$ which denotes the pilot symbol part of $\mathbf{s}$ is given by
\begin{equation}\label{ST}
    \mathbf{s}_T = \{ \sum_{l=0}^{L-1} \beta \alpha_l e^{-j2\pi pK \Delta f \tau_l} + \mathbf{w}[pK] \}_{p=0}^{P-1}.
\end{equation}

Our objective is to minimize the normalized mean square error (NMSE) representing the channel estimation error, which is given by
\begin{equation}
{
\text{NMSE} = \frac{||\mathbf{h} - \hat{\mathbf{h}}||_2^2}{||{\mathbf{h}}||_2^2},}
\end{equation}
where $\hat{\mathbf{h}} \in \mathbb{C}^N$ denotes  the estimate of $\mathbf{h}$ and $||\cdot||_2$ is the operator of 2-norm.

\section{Method of GSD-SCE}

Our approach to the sparse channel estimation is to directly extract the parameters of the $L$ multipaths. To achieve this, we need to characterize the excess delay of each path in a tractable manner.

Let us have a close look at the $l$-th path of $\mathbf{s}_T$. We momentarily neglect the noise for the brevity of the explanation. Then, recalling \eqref{ST}, the sequence of the $l$-th path within $\mathbf{s}_T$ is as follows:
\begin{equation} \label{lth_path}
    \beta \alpha_l \{ 1, e^{-j2\pi K \Delta f \tau_l}, e^{-j2\pi 2K \Delta f \tau_l}, e^{-j2\pi 3K \Delta f \tau_l}, \cdots \}.
\end{equation}
Observe that \eqref{lth_path} is a geometric sequence with the initial term of $\beta \alpha_l$ and the common ratio of $e^{-j2\pi K \Delta f \tau_l}$. This implies that $\mathbf{s}_T$ can be regarded as a non-orthogonal superposition of $L$ geometric sequences.

A geometric sequence is characterized by two parameters: initial term and common ratio. Let $\mathbf{a}$ and $\mathbf{r}$ be the vectors of initial terms and common ratios of the geometric sequences that consist of $\mathbf{s}_T$. In other words,
\begin{equation}
\begin{split}
\mathbf{a} & = [\beta \alpha_0 , \cdots , \beta \alpha_{L-1}], \\
\mathbf{r} & = [e^{-j2\pi K \Delta f \tau_0},\cdots,e^{-j2\pi K \Delta f \tau_{L-1}}].
\end{split}
\end{equation}
The entire CFR for $N$ subcarriers, $\mathbf{h}$, can be reconstructed by obtaining $2L+1$ unknowns: $\mathbf{a}$, $\mathbf{r}$, and $L$.

We employ GSD-ST, which is a mathematical technique of handling geometric sequences\cite{lee2019geometric}.
The fundamental idea of GSD-ST is to transform a sequence ($\mathbf{s}_T$ for our case) to a virtual $L$-dimensional space and exploit its geometric properties. The concept of $L$-simplex\footnote{{$L$-simplex is defined as an $L$-dimensional polytope which is the convex hull of its $L + 1$ $L$-vertices \cite{grunbaum1969convex}.}} is utilized for the transformation. The iterative truncated singular value decomposition (SVD) is used for the implementation of GSD-ST to mitigate the impact of noise. Following the process of GSD-ST, the proposed GSD-SCE is divided into three phases. The subsequent sections explain these phases.

\subsection{Phase 1: obtaining $L$}
The first phase is devoted to obtaining $L$. Let $\hat{L}$ be an estimate of $L$. Then, we create a new sequence which consists of the volumes of $\hat{L}$-simplexes out of $\mathbf{s}_T$. According to Theorem 1 in \cite{lee2019geometric}, such a sequence becomes a non-zero geometric sequence if and only if $\hat{L}$ is the correct estimate of $L$.

Firstly, let us make $\hat{L}$-dimensional vertices out of $\mathbf{s}_T$. The $k$-th vertex is denoted by $\mathbf{v}_k \in \mathbb{C}^{\hat{L}}$ and is defined as follows:
\begin{equation}\label{v_k}
    \mathbf{v}_k = [\mathbf{s}_T(k),\cdots,\mathbf{s}_T(k+\hat{L}-1)]^T,
\end{equation}
where $(\cdot)^T$ is the transpose of an input.

Secondly, let $\chi_{\hat{L}}(\mathbf{v}_k,\cdots,\mathbf{v}_{k+\hat{L}-1})$ be the $\hat{L}$-simplex by connecting the origin point and given  $\hat{L}$ consecutive vertices. In addition, let $\Lambda(\chi_{\hat{L}}(\mathbf{v}_k,\cdots,\mathbf{v}_{k+\hat{L}-1}))$ denote the volume of the simplex, i.e.,
\begin{equation}\label{vol}
\Lambda(\chi_{\hat{L}}(\mathbf{v}_k,\cdots,\mathbf{v}_{k+\hat{L}-1})) = \frac{\mathrm{det}([\mathbf{v}_k,\cdots,\mathbf{v}_{k+\hat{L}-1}])}{\hat{L}!},
\end{equation}
where $\mathrm{det}(\cdot)$ is the determinant of an input.

Thirdly, we define $\Omega_{\hat{L}}$ as the series of the volumes of $\hat{L}$-simplexes, i.e.,
\begin{equation}\label{omega}
    \Omega_{\hat{L}} = \{ \Lambda ( \chi_{\hat{L}}(\mathbf{v}_k,\cdots,\mathbf{v}_{k+\hat{L}-1}) \}_{k=0}^{P-\hat{L}}.
\end{equation}
As stated before, $\Omega_{\hat{L}}$ becomes a non-zero geometric sequence if and only if $\hat{L} = L$. Therefore, we can obtain $L$ by examining whether \eqref{omega} is a geometric sequence for different $\hat{L}$ values.

\subsection{Phase 2: obtaining  $\mathbf{a}$ and $\mathbf{r}$}
In the second phase, we obtain $\mathbf{a}$ and $\mathbf{r}$ with the information of known $L$. The essence of this phase to create another new series of $L$-simplexes from $2L$ samples of $\mathbf{s}_T$. According to Theorem 2 in \cite{lee2019geometric}, we can construct an $L$-th order polynomial equation whose roots are equal to $\mathbf{r}$.

To create the series, firstly compose $L+1$ consecutive vertices, $\mathbf{v}_0,\cdots,\mathbf{v}_{L}$, as in \eqref{v_k}.
Next, let $\aleph$ be the series that we will utilize. Its length is $L+1$, and it is formed by the lexicographical combination of the vertices as follows:
\begin{equation}
    \begin{split}
        \aleph & = \{ \chi_L (\mathbf{v}_0, \mathbf{v}_1, \cdots, \mathbf{v}_{L-1}), \chi_L (\mathbf{v}_0, \mathbf{v}_1, \cdots, \mathbf{v}_{L}), \\ & \cdots , \chi_L (\mathbf{v}_0, \mathbf{v}_2, \cdots, \mathbf{v}_{L}), \chi_L (\mathbf{v}_1, \mathbf{v}_2, \cdots, \mathbf{v}_{L}) \}.
    \end{split}
\end{equation}
As mentioned above, $\mathbf{r}$ is equal to the roots of the following $L$-th order polynomial equation:
\begin{equation}\label{g}
    \sum_{k=0}^{L} \Lambda(\aleph[k]) (-r)^{L-k} = 0.
\end{equation}

Once $\mathbf{r}$ is obtained, $\mathbf{a}$ can be found conveniently by the following matrix pseudo-inversion:
\begin{equation}
   \mathbf{a}=\mathbf{R}^+ \mathbf{s}_T,
\end{equation}
where $\mathbf{R} \in \mathbb{C}^{L \times P}$ is the matrix constructed by $\mathbf{r}$ satisfying $\mathbf{R}[i,j]: = (\mathbf{r}[i])^j$ for $i,j = 0,1,2,\cdots$, and $(\cdot)^+$ is the pseudo-inverse operation.

\subsection{Phase 3: extracting channel parameters}

The extraction of the channel parameters $\alpha_0,\cdots,\alpha_{L-1}$ and $\tau_0,\cdots,\tau_{L-1}$ is straightforward from $\mathbf{a}$, $\mathbf{r}$, and the predetermined values of $\beta$, $K$, and $\Delta f$. That is,
\begin{equation}\label{ch}
\hat{\alpha}_l \leftarrow \mathbf{a}[l]/\beta
\text{ and } \hat{\tau}_l \leftarrow \frac{-\angle \mathbf{r}[l]}{2\pi K \Delta f},
\end{equation}
where $\angle(\cdot)$ is the phase of a complex-valued input. Finally, $\hat{\mathbf{h}}$ can be reconstructed as follows:
\begin{equation}{\label{ds}}
    \hat{\mathbf{h}} = \{ \sum_{l=0}^{L-1} \hat{\alpha}_l e^{-j2\pi n \Delta f \hat{\tau}_l}\}_{n=0}^{N-1}.
\end{equation}

\begin{rem}
Under an ideal assumption of perfect pilot estimation, GSD-SCE can lead to the error-free channel reconstruction of the whole bandwidth. In this case, the only source of the error in the GSD-SCE scheme is phase ambiguity as indicated in \eqref{ch}. It occurs when the excess delay happens to be large such that $\angle \mathbf{r}[l]$ stretches out of the range of $[-2\pi,0)$. Therefore, the error-free channel estimation is achieved if the following inequality is satisfied:
\begin{equation}\label{con}
    \tau_{max} < \frac{1}{K \Delta f},
\end{equation}
where $\tau_{max}$ is the upper-bound of $\tau$.
\end{rem}

\subsection{{Complexity analysis}}
The computational complexity of GSD-SCE is $\mathcal{O}((L+1)L^{2.373})$, which is equivalent to obtaining $\{\det \aleph [j]\}_{j=0}^{L}$ if $L \leq 4$ \cite{aho1974design}.
Otherwise, eigenvalue decomposition of $L$-by-$L$ companion matrix is additionally required, i.e., eventually $\mathcal{O}((L+2)L^{2.373})$.
This is because the major computation of GSD-SCE is devoted to constructing and solving \eqref{g}, i.e., an $L$-th order polynomial equation. Therefore, the method is efficient when $L \leq 4$.

\section{Simulation Results}

The performance of the proposed GSD-SCE is evaluated via $5000$ Monte Carlo simulation experiments. For the simulation parameters, we set $\Delta f = 60$ kHz. We further assume that $\alpha$ follows $\mathcal{CN}(0,1)$ and $\tau$ follows $\text{Unif}(0,1\mu \text{sec})$. For a performance comparison, we select three schemes: MPM \cite{hua1990matrix} and two CS-based methods, i.e., atomic norm minimization (ANM) based continuous CS (CCS) \cite{pejoski2015estimation} and OMP based CS \cite{tropp2007signal} with a partial discrete Fourier transform (DFT) matrix whose number of bases is 5000.

\begin{figure}[t!]
\centering
\includegraphics[width=0.38\textwidth]{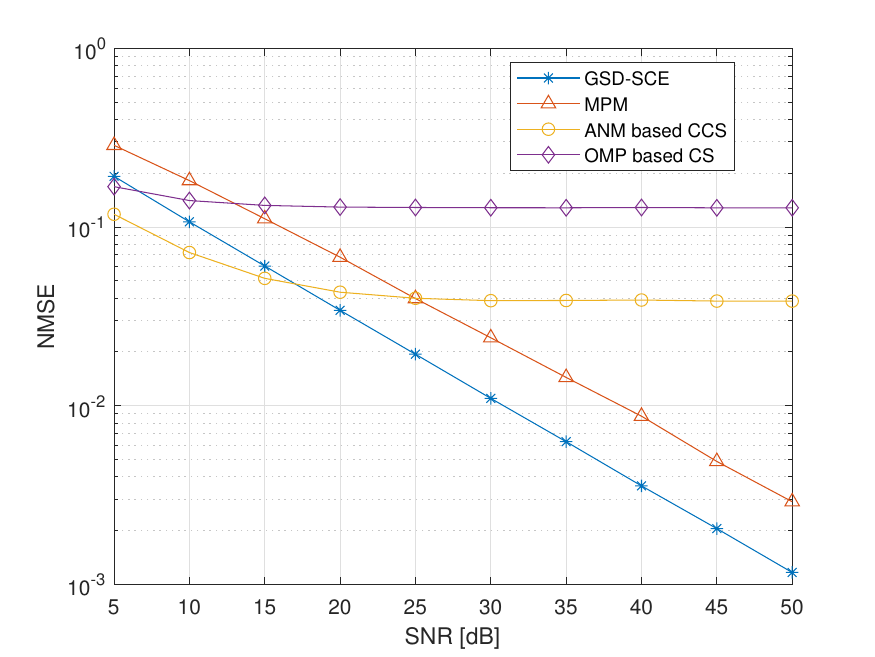}
\caption{NMSE according to SNR ($N$: 2048, $K$: 12, $L$: 4).}
\label{Fig_SNR}
\end{figure}

\begin{figure}[t!]
\centering
\includegraphics[width=0.38\textwidth]{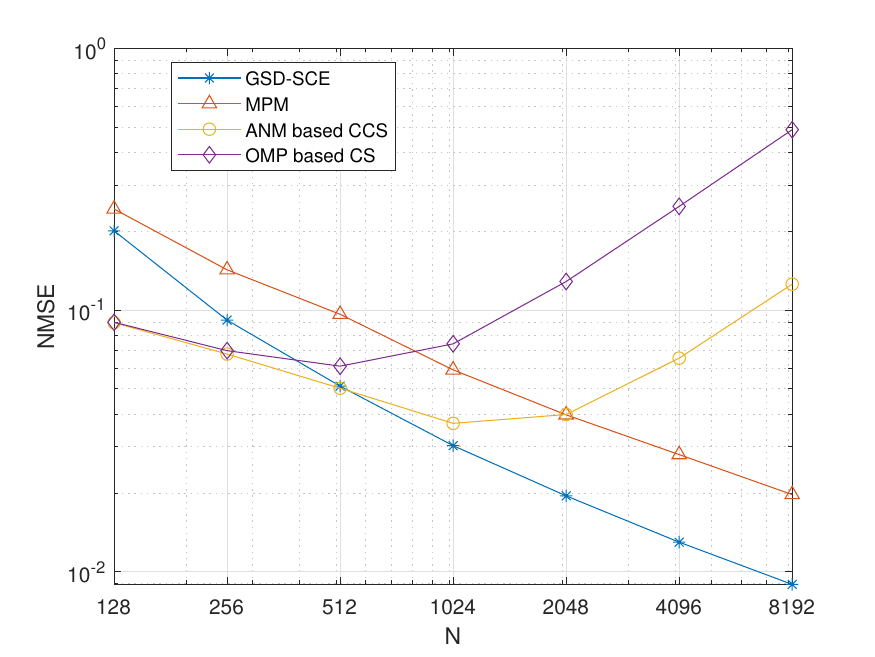}
\caption{NMSE according to $N$ (SNR: 25 dB, $K$: 12, $L$: 4).}
\label{Fig_N}
\end{figure}

\begin{figure}[t!]
\centering
\includegraphics[width=0.38\textwidth]{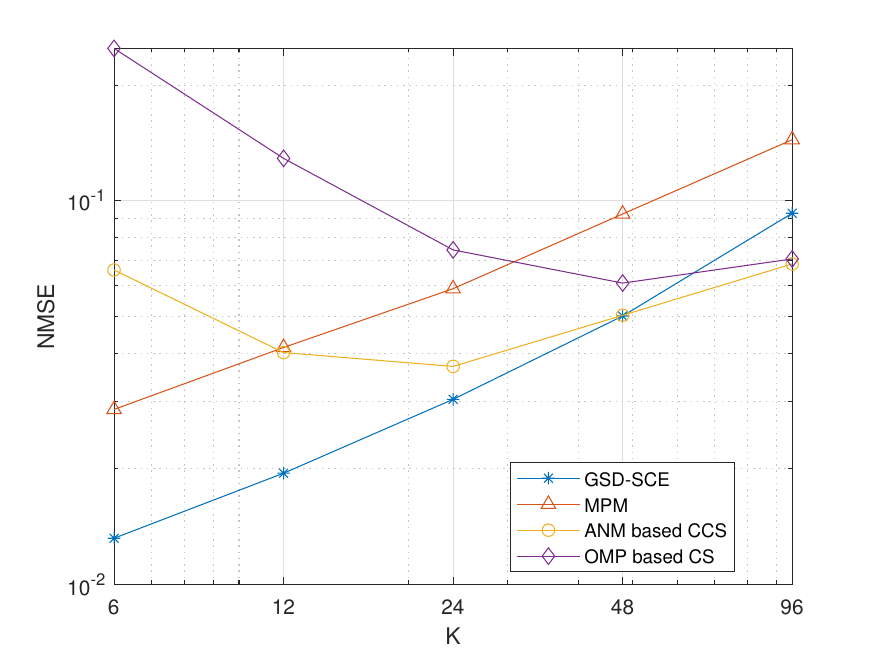}
\caption{NMSE according to $K$ (SNR: 25 dB, $N$: 2048, $L$: 4).}
\label{Fig_K}
\end{figure}

\begin{figure*}[t!]
     \centering
     \begin{subfigure}[b]{0.3\textwidth}
         \centering
         \includegraphics[width=\textwidth]{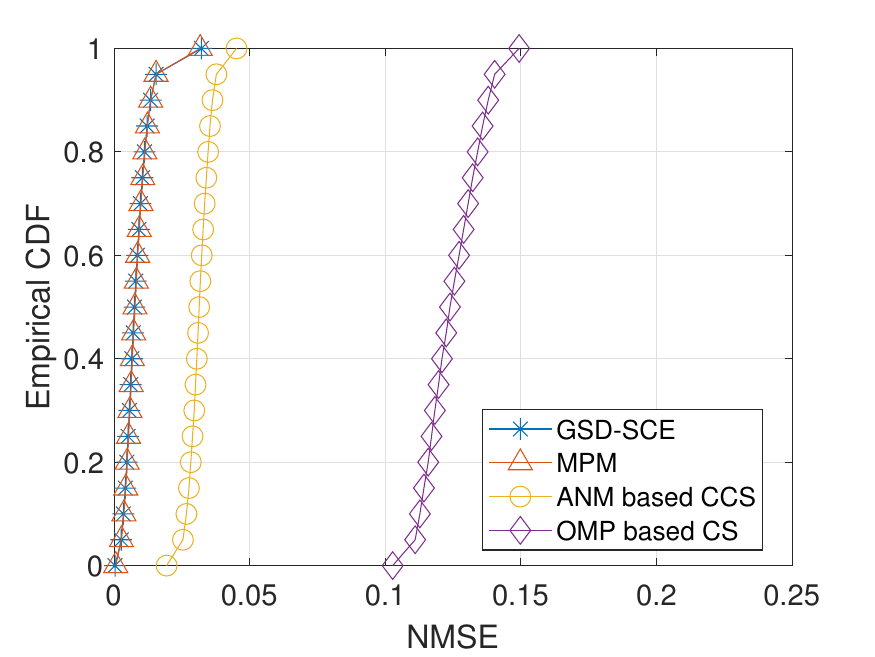}
         \caption{$L=1$}
         \label{fig:NE}
     \end{subfigure}
     \hfill
     \begin{subfigure}[b]{0.3\textwidth}
         \centering
         \includegraphics[width=\textwidth]{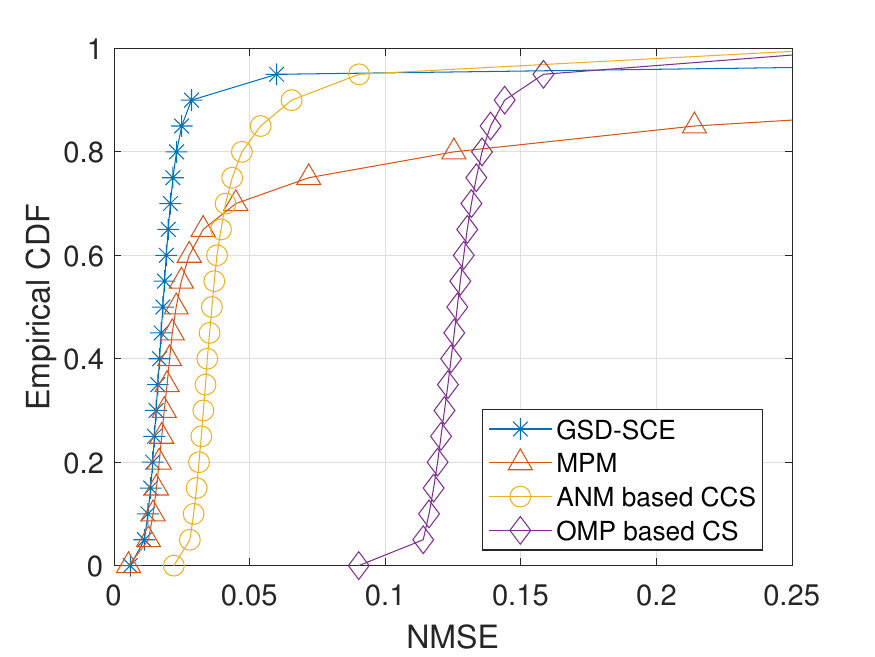}
         \caption{$L=4$}
         \label{fig:NP}
     \end{subfigure}
     \hfill
     \begin{subfigure}[b]{0.3\textwidth}
         \centering
         \includegraphics[width=\textwidth]{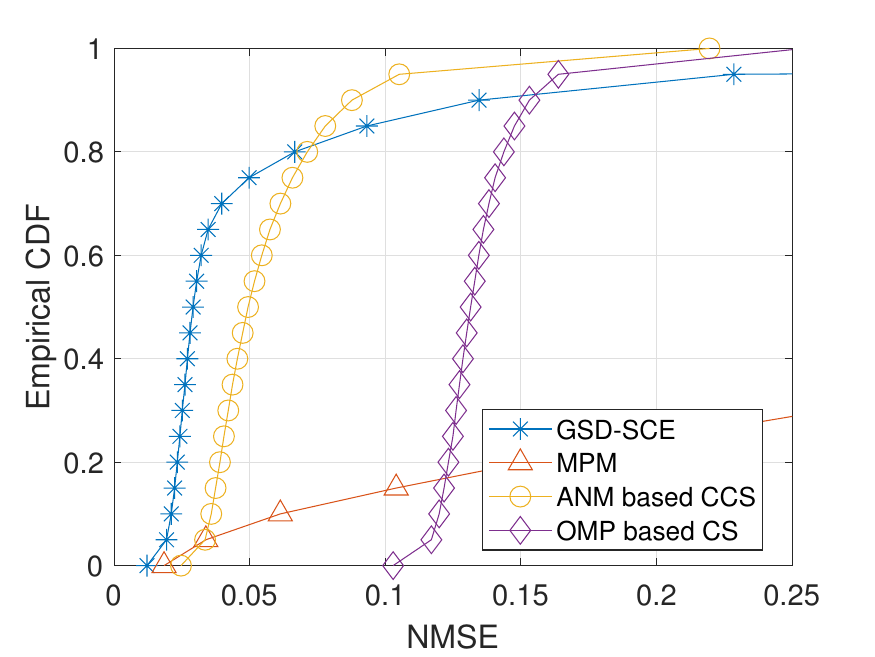}
         \caption{$L=8$}
         \label{fig:NQ}
     \end{subfigure}
        \caption{NMSE according to $L$ (SNR: 25 dB, $N$: 2048, $K$: 12).}
        \label{Fig_L}
\end{figure*}

The effect of SNR is illustrated in Fig. \ref{Fig_SNR}.
It is observed that the CS-based methods are more robust to noise in low SNR regime. However, GSD-SCE starts to outperform the CS-based methods at the SNR of about 15 dB, and the gap widens as SNR increases. The performances of the CS-based methods are saturated even in the high SNR regime. This is because the CS-based methods have a limitation in resolving the inter-path interference coming from the non-orthogonality among multipaths. In contrast, the performance of GSD-SCE decreases linearly as SNR increases in the log-log scale. The MPM method exhibits a similar behavior as the proposed GSD-SCE. However, GSD-SCE is superior to MPM in the whole SNR range. Therefore, we can conclude that GSD-SCE gives a better NMSE performance than MPM under practical SNR conditions although they show similar trends.

In Fig. \ref{Fig_N}, we examine the effect of the number of subcarriers, i.e., $N$. It is noteworthy that the NMSE of GSD-SCE improves as $N$ increases. This suggests that the performance of GSD-SCE is not strongly affected by the number of data symbols, $(N-P)$. Rather, its performance depends on the ratio of $P$ and $L$. This is because GSD-SCE interprets the observed signal as a superposition of geometric sequences and directly extracts $2L$ unknowns. Therefore, GSD-SCE is advantageous to wide bandwidth. As for MPM, similar to Fig. \ref{Fig_SNR}, it shows the same trend but inferior performance to GSD-SCE. On the other hand, the CS-based methods are more sensitive to the increase in $(N-P)$, resulting in a deteriorating performance.

The effect of the pilot spacing, i.e., $K$, is illustrated in Fig. \ref{Fig_K}. The performances of CS-based methods improve as $K$ increases in the beginning. This is because the probability of detecting the best correlated basis increases with larger $K$ owing to the decreased correlations among the multipaths. However, the number of pilots decreases at the same time, and thus the CS-based methods suffers a performance degradation after a certain $K$ value. The performance of GSD-SCE keeps worsening because the number of samples for denoising becomes insufficient. Nonetheless, GSD-SCE shows a superior performance up to $K = 48$.

In Fig. \ref{Fig_L}, we examine the effect of the number of multipaths, i.e., $L$. Although GSD-SCE generally performs better than other methods, it is sensitive to the increase in $L$. When $L=8$, a fraction of poor channel estimation results is observed even with the excellent average performance. This is because numerical errors may occur when \eqref{g} is solved. A closed form solution for \eqref{g} exists for lower values of $L$ (such as $L \leq 4$), whereas the performance of root-finding algorithms play an important role in the accuracy of GSD-SCE when $L>4$. This suggests that GSD-SCE functions better under a sparser channel.

\section{Conclusion}

We introduced a new method for the sparse channel estimation, which we term GSD-SCE.
Different from the conventional CS-based methods, our approach is to directly obtain the channel parameters of multipaths. To achieve this, we utilized the fact that the $L$-tapped CIR in time domain can be represented as the superposition of $L$ geometric sequences in frequency domain. Then, we employed recently developed GSD-ST method to extract the parameters of the geometric sequences. Performance comparison with existing methods shows that GSD-SCE is advantageous in higher SNR, wider bandwidth, and sparser channel conditions. Improving the performance of GSD-SCE under low SNR regime remains as further research.

\bibliographystyle{IEEEtran}
\bibliography{references}

\end{document}